\documentclass[sigconf]{acmart}

\usepackage{booktabs} 
\usepackage{subfigure}
\usepackage{multirow}

\newcommand{\para}[1]{\vspace{0.01in}\noindent\textbf{#1 }}

\copyrightyear{2019}
\acmYear{2019} 
\setcopyright{iw3c2w3}
\acmConference[WWW '19]{Proceedings of the 2019 World Wide Web Conference}{May 13--17, 2019}{San Francisco, CA, USA}
\acmBooktitle{Proceedings of the 2019 World Wide Web Conference (WWW '19), May 13--17, 2019, San Francisco, CA, USA}
\acmPrice{}
\acmDOI{10.1145/3308558.3313705}
\acmISBN{978-1-4503-6674-8/19/05}

\usepackage{balance}

\begin{document}

\title{Unifying Knowledge Graph Learning and Recommendation: \\ Towards a Better Understanding of User Preferences}

\author{Yixin Cao}
\affiliation{National University of Singapore}
\email{caoyixin2011@gmail.com}

\author{Xiang Wang}
\affiliation{National University of Singapore}
\email{xiangwang1223@gmail.com}

\author{Xiangnan He}
\authornote{Corresponding author.}
\affiliation{University of Science and Technology of China}
\email{xiangnanhe@gmail.com}

\author{Zikun Hu}
\affiliation{National University of Singapore}
\email{zikunhu1016@gmail.com}

\author{Tat-Seng Chua}
\affiliation{National University of Singapore}
\email{dcscts@nus.edu.sg}

\renewcommand{\shortauthors}{Y. Cao et al.}

\begin{abstract}
  Incorporating knowledge graph (KG) into recommender system is promising in improving the recommendation accuracy and explainability. However, existing methods largely assume that a KG is complete and simply transfer the "knowledge" in KG at the shallow level of entity raw data or embeddings. This may lead to suboptimal performance, since a practical KG can hardly be complete, and it is common that a KG has missing facts, relations, and entities. Thus, we argue that it is crucial to consider the incomplete nature of KG when incorporating it into recommender system. 

  In this paper, we jointly learn the model of recommendation and knowledge graph completion. Distinct from previous KG-based recommendation methods, we transfer the \textbf{relation} information in KG, so as to understand the reasons that a user likes an item. As an example, if a user has watched several movies \textit{directed by} (relation) the \textit{same person} (entity), we can infer that the director relation plays a critical role when the user makes the decision, thus help to understand the user's preference at a finer granularity. 
  
  Technically, we contribute a new translation-based recommendation model, which specially accounts for various preferences in translating a user to an item, and then jointly train it with a KG completion model by combining several transfer schemes. Extensive experiments on two benchmark datasets show that our method outperforms state-of-the-art KG-based recommendation methods. Further analysis verifies the positive effect of joint training on both tasks of recommendation and KG completion, and the advantage of our model in understanding user preference. We publish our project at \url{https://github.com/TaoMiner/joint-kg-recommender}.
\end{abstract}

%
%

\ccsdesc[500]{Information systems~Personalization}

\keywords{Item Recommendation;Knowledge Graph;Embedding;Joint Model;}

\maketitle

\section{Introduction}
Knowledge Graph (KG) is a heterogeneous structure that stores the world's knowledge in the form of machine-readable graphs, where nodes denote entities, and edges denote the relations between entities. Since proposed, KG has attracted much attention in many fields, ranging from recommendation~\cite{xiang2018reasoning}, dialogue system~\cite{DBLP:conf/acl/KanHLJRY18,DBLP:conf/cikm/JinLRCLZY18}, to information extraction~\cite{cao2018neural}. Focusing on recommendation, the structural knowledge has shown great potential in providing rich information about the items, offers a promising solution to improving the accuracy and explainability of recommender systems.

\begin{figure}[t]
  \centerline{\includegraphics[width=0.47\textwidth]{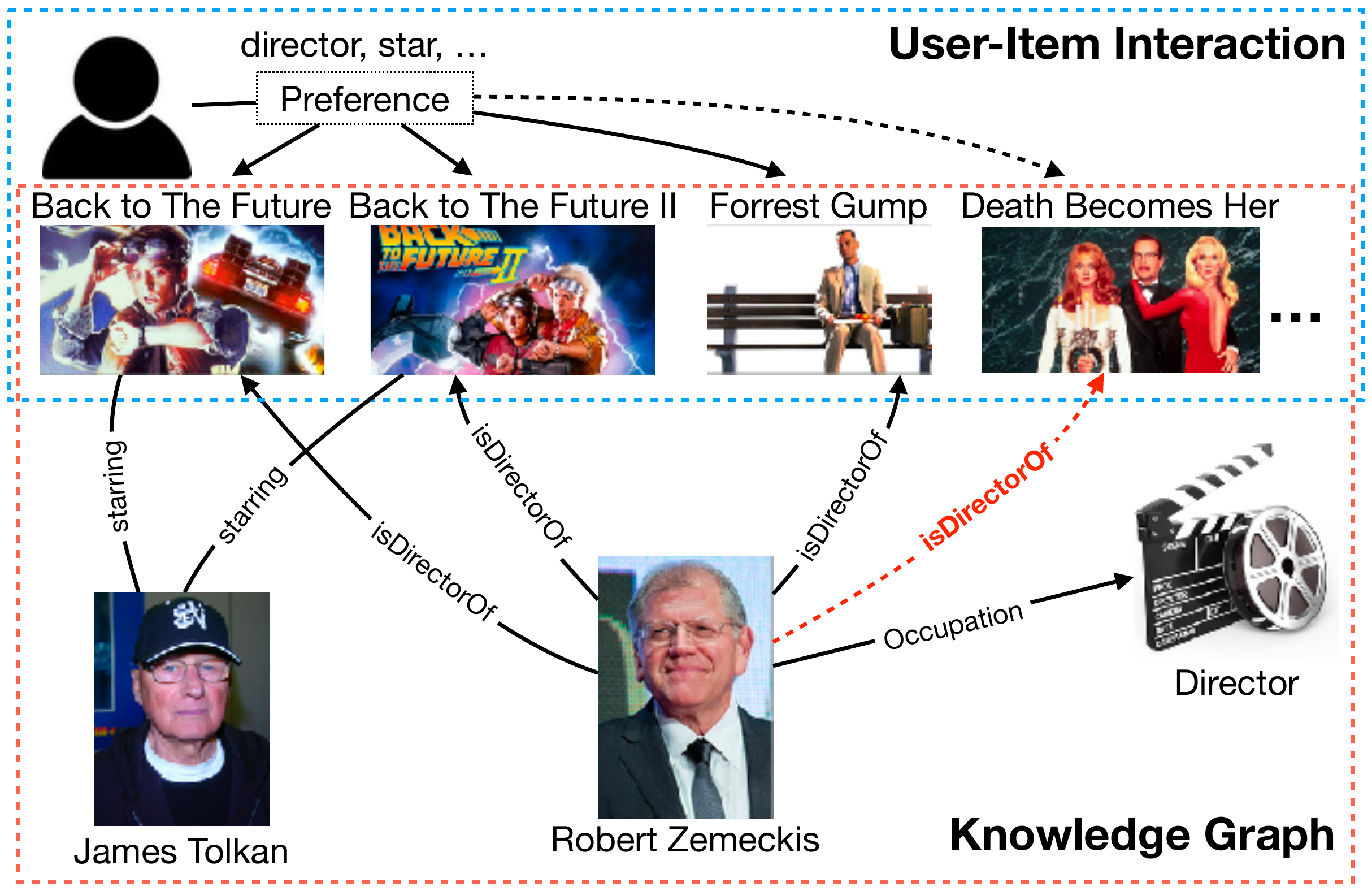}}
  \vspace{-0.4cm}
  \caption{An illustrative example on the necessity of considering the missing relations in KG for recommendation.}
  \vspace{-0.4cm}
  \label{fig:example}
\end{figure}

Nevertheless, existing KGs (e.g., DBPedia~\cite{lehmann2015dbpedia}) are far from complete, which limits the benefits of the transferred knowledge. As illustrated in Figure~\ref{fig:example}, the red dashed line between \textit{Robert Zemeckis} and \textit{Death Becomes Her} indicates a missing relation \textit{isDirectorOf}. Assuming that the user has chosen movies \textit{Back to The Future I \& II} and \textit{Forrest Gump}, by using the KG, we can attribute the reason of user's choices to the director \textit{Robert Zemeckis}. In this case, although we have accurately captured the user's preference on movies, we may still fail to recommend \textit{Death Becomes Her} (which is also of interest to the user), due to the missing relation in the KG (cf. the red dashed line). As such, we believe that it is critical to consider the incomplete nature of KG when using it for recommendation, and more interestingly, can the completion of KG benefit from the improved modeling of user-item interactions?

In this paper, we propose to unify the two tasks of recommendation and KG completion in a joint model for mutually enhancements. The basic idea is twofold: 1) utilizing the facts in KG as auxiliary data to augment the modeling of user-item interaction, and 2) completing the missing facts in KG based on the enhanced user-item modeling. For example, we are able to understand the user's preference on director via the related entities and relations; meanwhile we can predict that \textit{Robert Zemeckis} is the director of \textit{Death Becomes Her}, if there exist some users who like the movie also have a preference of other movies directed by \textit{Robert Zemeckis}.

Although many prior efforts have leveraged KG in recommender systems~\cite{piao2017factorization,DBLP:journals/corr/abs-1803-06540,CKE,DBLP:conf/sigir/HuangZDWC18,DBLP:conf/pakdd/ZhouLXXYLZ18,DBLP:conf/esws/PiaoB18b}, there are few works that jointly model the two tasks of knowledge graph learning and recommendation. CoFM~\cite{DBLP:conf/esws/PiaoB18} is the most similar work that aligns the two latent vector spaces in each task together by regularizing or sharing entity and item embeddings if they refer to the same thing. However, it ignores the important role of entity relations in user-item modeling, and fails to offer interpretation ability.

In this work, we propose a \textbf{T}ranslation-based \textbf{U}ser \textbf{P}reference model (TUP) to integrate with KG learning seamlessly. The key idea is that there exists multiple (implicit) relations between users and items, which reveal the preferences (i.e., reasons) of users on consuming items. An example of the ``preference'' is the director information in Figure 1 that drives the user to watch the movies \textit{Back to Future I \& II} and \textit{Forest Gump}. 
While we can pre-define the number of preferences and train TUP from user-item interaction data, the preferences are expressed as latent vectors that are opaque to deeper understanding. 
To endow the preferences with explicit semantics, we align them with the relations in KG, capturing the intuition that the type of item attributes plays a crucial role in user decision-making process. Technically speaking, we transfer the relation embeddings as well as entity embeddings learned from KG to TUP, simultaneously training the KG completion and recommendation tasks. We term the method as \textbf{K}nowledge-enhanced TUP (KTUP) that jointly learns the representations of users, items, entities, and relations. The main contributions are summarized:
\begin{itemize}
    \item We propose a new translation-based model, which exploits the implicit preference representations to capture the relations between users and items.
    
    \item We emphasize the importance of jointly modeling item recommendation and KG completion to couple the preference representations with the knowledge-aware relations, thus empowering the model with explainability.
    
    \item We perform extensive experiments on two datasets for top-$N$ recommendation and KG completion tasks, which verify the rationality of joint learning.
    Experimental results demonstrate the effectiveness and explainability of our model.
    
\end{itemize}

\section{Related Work}
Our proposed methods involve two tasks: item recommendation and KG completion. In this section, we first introduce related work for each task, before discussing their relations.

\subsection{Item Recommendation}
In the early stage of item recommendation, researchers focus on recommending similar users or items to a target user using history interactions alone, such as collaborative filtering (CF)~\cite{sarwar2001item}, factorization machines~\cite{rendle2010factorization}, matrix factorization techniques~\cite{koren2009matrix}, BPRMF~\cite{rendle2009bpr}. The key challenge here lies in extracting features of users and items to compute their similarity, namely \textbf{similarity-based methods}.

With the surge of neural network (NN) models, a host of methods extend \textbf{similarity-based methods with NN} and present a more effective mechanism in automatically extracting latent features of users and items for recommendation~\cite{he2017ncf,he2017nfm,he2018adversarial,chen2018sequential}. However, they still suffer from the data sparsity issue and cold-start problem. \textbf{Content-based methods} deal with the issues by introducing various side information, such as the contextual reviews~\cite{cheng2018aspect,mcauley2013hidden}, relational data~\cite{sharma2013social,feng2018temporal} and knowledge graphs~\cite{catherine2016personalized}. Another advantage of additional contents is the improved explainable ability to understand why an item is being recommended out of others. This has been found to be important for the effectiveness, efficiency, persuasiveness, and user satisfaction of recommender systems~\cite{zhang2014explicit,chen2018visually,wang2018tem}.

Among the side information, knowledge graphs (e.g., DBPedia~\cite{lehmann2015dbpedia}) show great potentials on recommendations due to its well-defined structures and adequant resources. This type of methods mostly transfer structural knowledge of entities from KG to user-item interaction modeling based on the given mapping between entities and items. We roughly classify them into two groups: the methods augmenting the data of user-item pairs with KG triplets, and the methods combining item and entity embeddings learned from different sources. In the first group, \citet{piao2017factorization} extracted lightweight features drived from KG (i.e. property-object, subject-property) for factorization machines. \citet{DBLP:journals/corr/abs-1803-06540} constructed a unified graph by adding a \textit{buy} relation between users and items, then applied transE~\cite{bordes2013translating} to model relational data. On the other hand, the methods in the second group usually improve the quality of item embeddings using entity embeddings if they refer to the same thing~\cite{CKE,DBLP:conf/sigir/HuangZDWC18}. \citet{DBLP:conf/esws/PiaoB18b} summerized the recommendation results using different entity embeddings (i.e. node2vec, doc2vec and transE), and found that node2vec improves the most. CoFM~\cite{DBLP:conf/esws/PiaoB18} first takes into account the refinements of entity embeddings from user-item modeling as another transfering task. However, the above methods heavily rely on the alignments between items and entities. \citet{DBLP:conf/pakdd/ZhouLXXYLZ18} introduced entity concepts in KG to deal with the sparisity issue of the alignments, but still fail to consider the importance of entity relations in transfering knowledge from KG.

Another line of work is \textbf{translation-based recommendation}, inspired by KG representation learning. It assumes that the selection of items satisfies translational relations in the latent vector space, where the relations are either regarded as related to users in sequential recommendation~\cite{he2017translation}, or modeled implicitly via Memory-based Attention~\cite{tay2018latent}. We thus improve this type of method by considering the N-to-N issue\footnote{N-to-N issue here means that one user may like multiple items, and also, several users may like a single item, which will be detailed in Section~\ref{sec:ht}.} in modeling user preferences as translational relations, which shall be further enhanced by transfering knowledge of entities and their relations from KG.

\subsection{KG Completion}
External knowledge has been found to be effective in many tasks of natural language processing, such as the question answering~\cite{zhang2015target}, which accelerates the popularity of KGs. Although there are a host of methods for finding entities~\cite{cao2017bridge,cao2018joint} and their relations from texts~\cite{lin2015modeling}, existing KGs are far from complete. Recent studies for KG completion show a great enthusiasm for learning low-dimensional representions of entities and relations while perserving structural knowledge of the graph. We roughly categorize such representation learning methods into two groups: translational distance models and semantic matching models.

TransE~\cite{bordes2013translating} first proposed the core idea of translational distance models that the relationship between two entities corresponds to a translation in their vector space. Although it is simple and effective, it is sometimes confusing because some relations can translate one entity to various entities, namely the 1-to-N problem. Similarly, there are other N-to-1 and N-to-N problems. To address these problems, a host of methods extended TransE by introducing additional hyperplanes~\cite{wang2014knowledge}, vector spaces~\cite{lin2015learning}, textual information~\cite{wang2016text} and relational path~\cite{lin2015modeling}.

The second group measures the plausibility of facts by matching semantic representations of entities and relations through similarity-based scoring functions. RESCAL~\cite{nickel2012factorizing} represents each relation as a matrix to capture the compositional semantics between entities, and utilizes a bilinear function as similarity metrics. To simplify the learning of relation matrices, DistMult~\cite{yang2014embedding} restricts them to be diagonal, HolE~\cite{nickel2016holographic} defines a circular correlation~\cite{plate1995holographic} to compress relation matrices as vectors, and ComplEx~\cite{trouillon2016complex} introduces complex-value for asymmetric relations. Instead of modeling the compositional relation, another line of methods directly introduce NNs for matching. SME~\cite{bordes2014semantic} learns relation specific layers for head entity and tail entity, respectively, and then feeds them into the final matching layer (e.g., dot production), while NAM~\cite{liu2016probabilistic} conducts the semantic matching with a deep architecture.

\subsection{Relationship between Two Tasks}
Items usually correspond to entities in many fields, such as books, movies and musics, making it possible to transfer knowledge between fields. These information involving in two tasks are complementary, revealing the connectivity among items or between users and items. In terms of models, the two tasks both aim to rank candidates given a query (i.e., an entity or a user) as well as their implict or explict relatedness. For example, KG completion aims to find correct movies (e.g., \textit{Death Becomes Her}) for the person \textit{Robert Zemeckis} given the explicit relation \textit{isDirectorOf}, while item recommendation aims at recommending movies for a target user satisfying some implicit preference. Therefore, we are able to fill in the gap between item recommendation and KG completion via a joint model, to systematically investigate how the two tasks impact each other.

\section{Preliminary}
Before introducing our proposed methods, let's first formally define the two tasks as well as TransH~\cite{wang2014knowledge} as the component for KG completion in our model.

\subsection{Tasks and Notations}

\para{Item Recommendation}: Given a list of user-item interactions $\mathcal{Y}=\{(u, i)\}$, we use implicit feedback as the protocol so that each pair $(u, i)$ implies the user $u\in\mathcal{U}$ consumes the item $i\in\mathcal{I}$. The goal is to recommend top-$N$ items for a target user.

\para{KG Completion}: A \textbf{Knowledge Graph} $\mathcal{KG}$ is a directed graph composed of \textit{subject}-\textit{property}-\textit{object} triple facts. Each triplet denotes that there is a relationship $r$ from head entity $e_h$ to tail entity $e_t$, formally defined by $(e_h, e_t, r)$, where $e_h,e_t\in\mathcal{E}$ are entities and $r\in\mathcal{R}$ are relations. Due to the incompleteness nature of KGs, KG completion is to predict the missing entity $e_h$ or $e_t$ for a triplet $(e_h, e_t, r)$, which can also be regarded as recommending top-N entities for a target $(e_t,r)$ or $(e_h,r)$.

\para{TUP} denotes the model for item recommendation. It takes a list of user-item pairs $\mathcal{Y}$ as input, and outputs a relevance score $g(u,i;p)$ indicating the likelihood that $u$ likes $i$, given the preference $p\in\mathcal{P}$, where the number of the preference set $\mathcal{P}$ is predefined. For each user-item pair, we induce a preference, serving as a similar role with the relation for two entities. To deal with the N-to-N issue, we introduce preference hyperplanes, and assign each preference with two vectors: $\mathbf{w}_p$ for the projection to a hyperplane, $\mathbf{p}$ for the translation between users and items.

\para{KTUP} is a multi-task architecture. Given $\mathcal{KG}$, $\mathcal{Y}$, and a set of item-entity alignments $\mathcal{A}=\{(i,e)|i\in\mathcal{I},e\in\mathcal{E}\}$, where each $(i,e)$ means that $i$ can be mapped to an entity $e$ in the given KG. KTUP is able to output not only $g(u,i;p)$, but also a score $f(e_h, e_t, r)$ indicating how possible the fact is true, based on the jointly learned embeddings of users $\mathbf{u}$, items $\mathbf{i}$, preferences $\mathbf{p},\mathbf{w}_p$, entities $\mathbf{e}$ and relations $\mathbf{r},\mathbf{w}_r$.

\begin{example}
  As shown in Figure~\ref{fig:example}, given a user, the interacted movies (e.g., \textit{Back to The Future I \& II} and \textit{Forrest Gump}), and the related triplets, KTUP is able to (1) figure out the user preference of the \textit{isDirectorOf} relation on movies, (2) recommend the movie \textit{Death Becomes Her} based on the induced preference, and (3) predict the missing head or tail entity for the triplet (\textit{Death Becomes Her}-\textit{isDirectorOf}-\textit{Robert Zemeckis}). The above three goals shall be achieved by considering not only the structural knowledge in KG but also the user-item interactions.
\end{example}

Next, we will briefly describe transH as the module of KG completion in our joint model.

\subsection{TransH for KG Completion}
Learning distributional representations of KG provides an effective and efficient way of manipulating entities while perserving their structural knowledge. TransE~\cite{bordes2013translating} is a widely used method due to its simplicity and remarkable effectiveness. Its basic idea is to learn embeddings for entities and relations, satisfying $\mathbf{e}_h+\mathbf{r}\approx \mathbf{e}_t$ if there is a triplet $(e_h, e_t, r)$ in KG. However, a single relation type may correspond to multiple head entities or tail entities, leading to serious 1-to-N, N-to-1 and N-to-N issues~\cite{wang2014knowledge}.

Therefore, TransH~\cite{wang2014knowledge} learns different representations for an entity conditioned on different relations. It assumes that each relation owns a hyperplane, and the translation between head entity and tail entity is valid only if they are projected to the same hyperplane. It defines an energy score function for a triplet as follows:

\begin{equation}
f(e_h, e_t, r)=\parallel \mathbf{e}_h^{\bot}+\mathbf{r}-\mathbf{e}_t^{\bot} \parallel
\end{equation}

\noindent where a lower score of $f(e_h, e_t, r)$ indicates that the triplet is possbily true, otherwise no. $\mathbf{e}_h^{\bot}$ and $\mathbf{e}_t^{\bot}$ are projected entity vectors:

\begin{equation}
\mathbf{e}_h^{\bot}=\mathbf{e}_h-\mathbf{w}_r^{\mathrm{T}} \mathbf{e}_h \mathbf{w}_r
\end{equation}

\begin{equation}
  \mathbf{e}_t^{\bot}=\mathbf{e}_t-\mathbf{w}_r^{\mathrm{T}} \mathbf{e}_t \mathbf{w}_r
\end{equation}

\noindent where $\mathbf{w}_r$ and $\mathbf{r}$ are two learned vectors of relation $r$, $\mathbf{w}_r$ denotes the projection vector of the corresponding hyperplane, and $\mathbf{r}$ is the translation vector. $\parallel \cdot \parallel$ denotes the L1-norm distance function used in the presented paper.

Finally, the training of TransH encourages the discrimination between valid triplets and incorrect ones using margin-based ranking loss:

\begin{equation}
  \mathcal{L}_{k}=\sum_{(e_h, e_t, r)\in\mathcal{KG}} \sum_{(e'_h, e'_t, r')\in\mathcal{KG}^-} [f(e_h, e_t, r)+\gamma-f(e'_h, e'_t, r')]_+
\end{equation}

\noindent where $[\cdot]_+\triangleq \max(0,\cdot)$, $\mathcal{KG}^-$ contains incorrect triplets constructed by replacing head entity or tail entity in a valid triplet randomly, and $\gamma$ controls the margin between positive and negative triplets.

\section{TUP for Item Recommendation}
Inspired by the above translation assumption between two entities in KG, we propose TUP to explictly models user preferences and regards them as translational relationships between users and items. Given a set of user-item interactions $\mathcal{Y}$, it automatically induces a preference for a user-items pair, and learns the embeddings of preference $\mathbf{p}$, user $\mathbf{u}$ and item $\mathbf{i}$, satisfying $\mathbf{u}+\mathbf{p}\approx \mathbf{i}$.

Considering the implicity and variety of user preferences, we design two main components in TUP: Preference Induction and Hyperplane-based Translation.

\subsection{Preference Induction}
\label{sec:pi}
Given a user-item pair $(u,i)$, this component is to induce a preference from a set of latent factors $\mathcal{P}$. These factors are shared by all users, and each $p\in\mathcal{P}$ denotes a different preference, which aims at capturing the commonality among users as global features complementary to user embeddings that focus on a single user locally. 
Similar with topic models, the number $P=|\mathcal{P}|$ is a hyperparameter, and we cannot nominate the exact meaning of each preference. 
With the help of KG, the number of preferences can be automatically set and each preference is assigned with explanations (Section~\ref{sec:joint}).

We design two strategies for preference induction: a hard approach that selects one out of the $P$ preferences, and a soft way that combines all preferences with attentions.

\subsubsection{Hard Strategy}
The intuition behind our hard strategy is that only a single preference takes effect when a user makes a decision over items. We use Straight-Through (ST) Gumbel SoftMax~\cite{jang2016categorical} for discretely sampling the preference given a user-item pair, which utilizes reparameterization trick for backpropagation, making it possible to calculate the continuous gradients of model parameters during the end-to-end training.

ST Gumbel SoftMax approximately samples one-hot vectors from a multi-classification distribution. Assuming that the probability of belonging to class $p$ in a $P$-way classification distribution is defined as log softmax:

\begin{equation}
  \phi(p)=\frac{\exp (\log (\pi_p))}{\sum_{j=1}^P \exp(\log(\pi_j))}
\end{equation}

\noindent where $\pi_p$ is the unnormalized output of a score function. Then, we sample a one-hot vector $\mathbf{z}=[z_1,\cdots,z_P]\in\mathbb{R}^P$ from the above distribution as follows:

\begin{equation}
  z_p=\left\{
               \begin{array}{lr}
               1, & p=\arg\max_j(\log(\pi_j)+g_j) \\
               0, & \text{otherwise}
               \end{array}
  \right.
\end{equation}

\noindent where $g=-\log(-\log(u))$ is the Gumbel noise and $u$ is generated by a certain noise distribution (e.g., $u\sim \mathcal{N}(0,1)$). The noise term increases the stochastic of $\arg\max$ function and makes the process become equivalent to drawing a sample accroding to a continuous probability distribution $\mathbf{y}=[y_1,\cdots,y_p,\cdots,y_P]$:

\begin{equation}
  y_p=\frac{\exp((\log(\pi_p)+g_p)/\tau)}{\sum_{j=1}^P \exp((\log(\pi_j)+g_j)/\tau)}
\end{equation}

This is called Gumbel-Softmax distribution, where $\tau$ is a temperature parameter. The related proofs can be found in the original papers.
  
Straight-Through (ST) gumbel-Softmax takes different paths in the forward and backward propagation, so as to maintain sparsity yet support stochastic gredient descent (SGD). In the forward pass, it discretizes the continuous probability distribution for a one-hot vector as mentioned above. And in the backward pass, it simply follows the continuous $\mathbf{y}$, thus the error signal is still able to be backpropagated.

In the hard strategy, we define the score function for $\pi_p$ as the similarity between the user-item pair and preference:

\begin{equation}
  \phi(u,i,p)=\text{Similarity}(\mathbf{u}+\mathbf{i}, \mathbf{p})
\end{equation}

We use dot product as similarity function.

\subsubsection{Soft Strategy}

Actually, a user might like an item according to various factors, which have no distinct boundary. Instead of selecting the most prominent preference, the soft strategy is to combine multiple preferences via the attention mechanism:

\begin{equation}
  \mathbf{p}=\sum_{p'\in\mathcal{P}}\alpha_{p'} \mathbf{p}'
\end{equation}

\noindent where $\alpha_{p'}$ is the attention weight of preference $p'$, and defined as proportional to the similarity score:

\begin{equation}
  \alpha_{p'}\propto \phi(u,i,p')
\end{equation}

\subsection{Hyperplane-based Translation}
\label{sec:ht}

\begin{figure}[htb]
  \centering
  \subfigure[TransE]{
  \label{fig:idea_e}\includegraphics[width=0.17\textwidth]{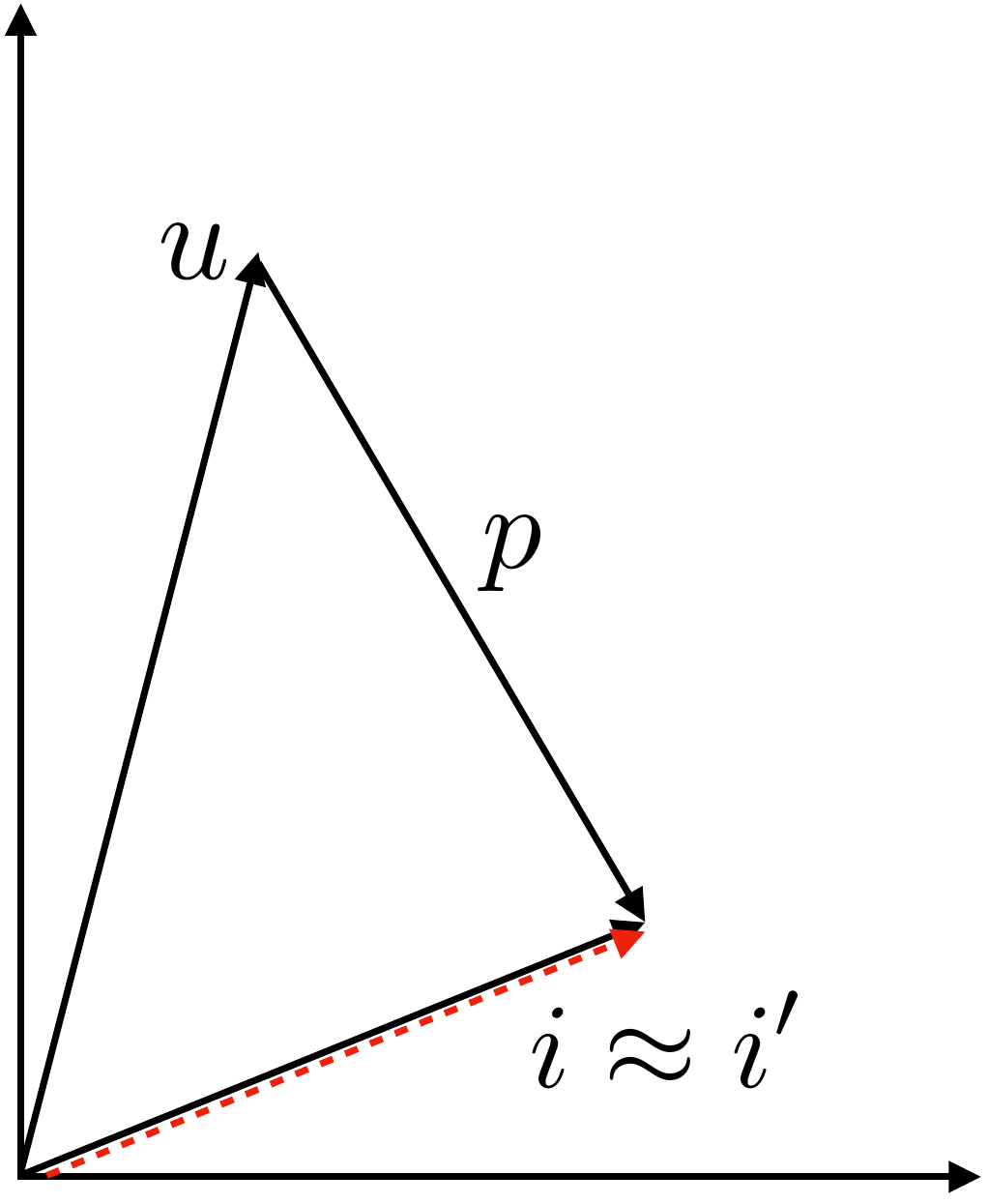}}
  \subfigure[TransH]{
  \label{fig:idea_h}\includegraphics[width=0.27\textwidth]{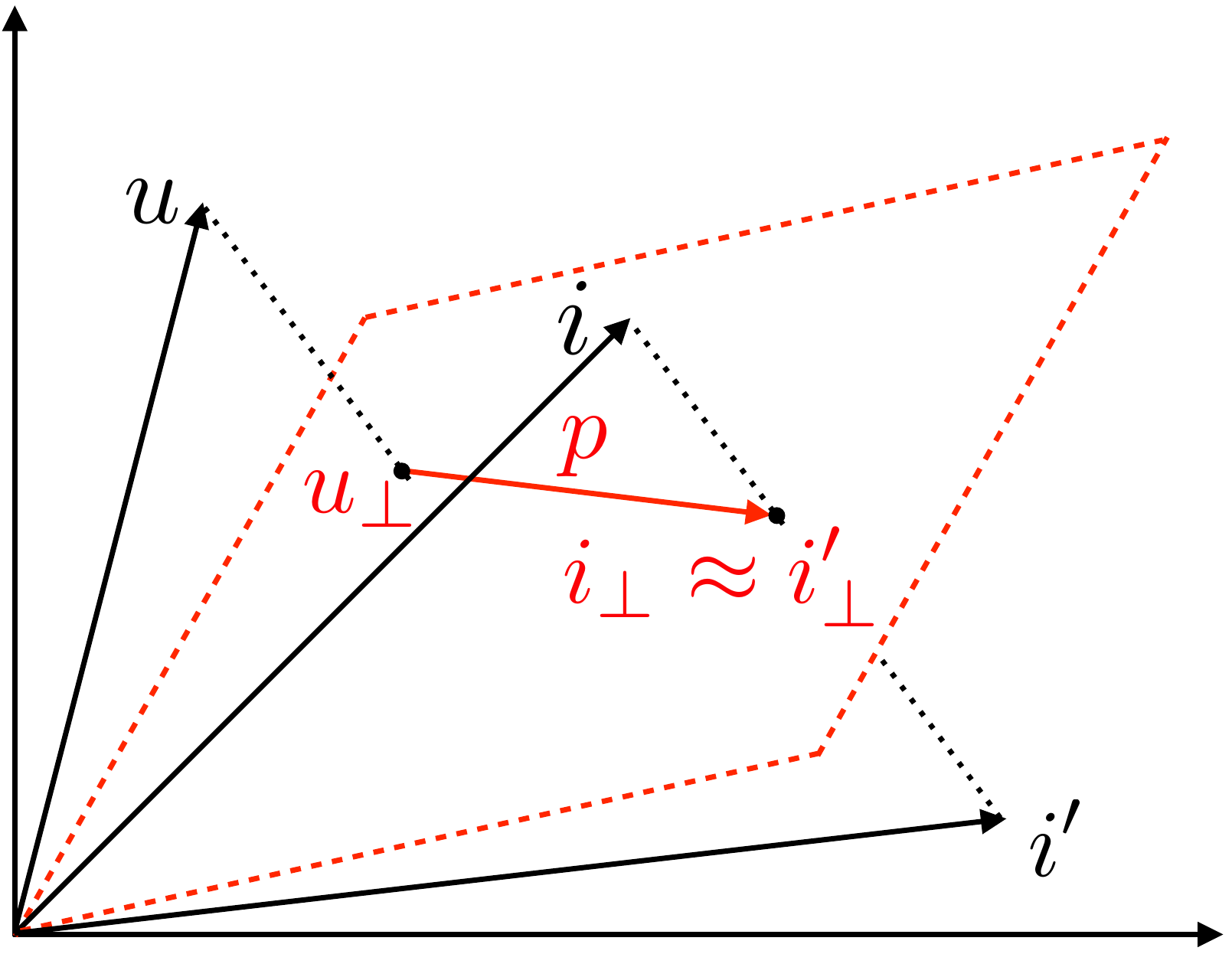}}
  \caption{Illustration of the two translation schemes for item recommendation}
  \label{fig:idea}
\end{figure}

\begin{figure*}[htb]
  \centerline{\includegraphics[width=0.73\textwidth]{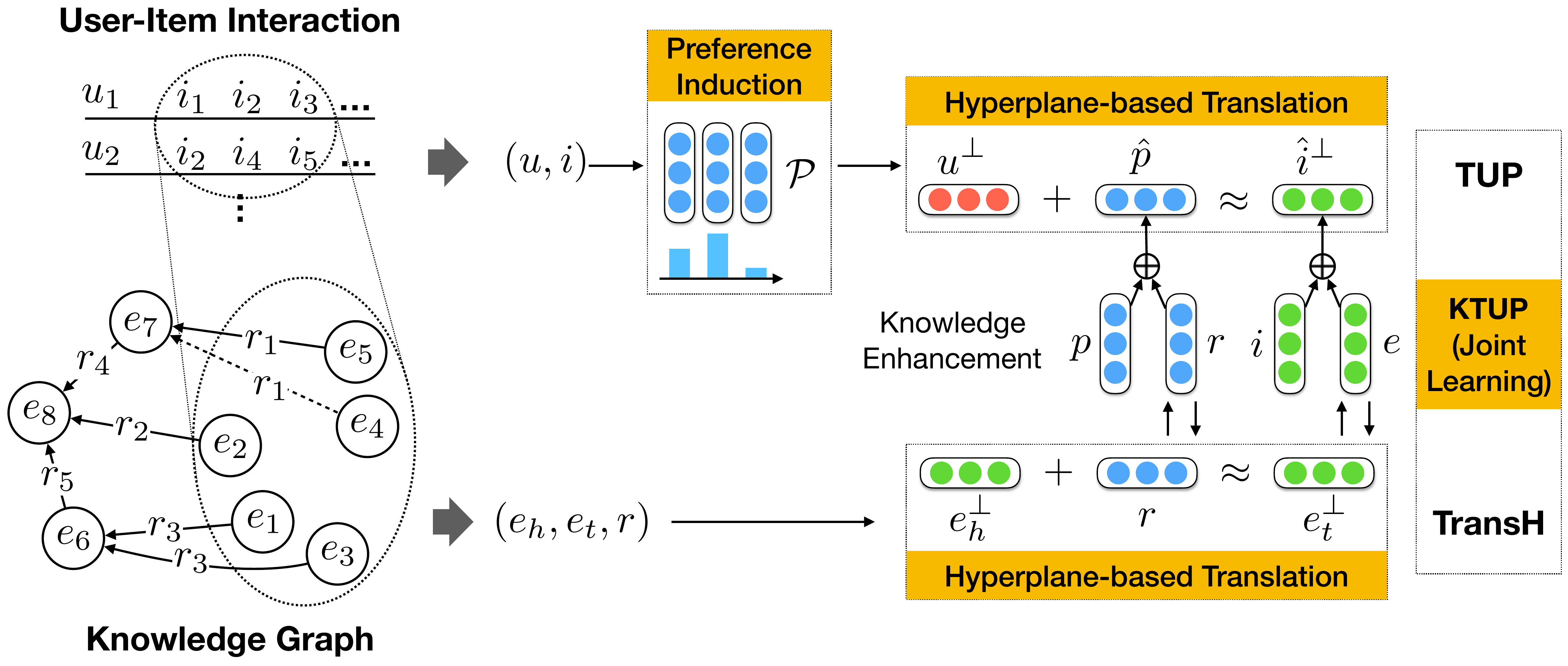}}
  \caption{Framwork of KTUP. At the top is TUP for item recommendation including two components: preference induction and hyperplane-based translation. KTUP jointly learns TUP and TransH to enhance the item and preference modeling by transfering knowledge of entities as well as relations.}
  \label{fig:framework}
\end{figure*}

Inspired by TransH, we introduce the hyperplane to deal with the variety of preferences. That is, different users may share the same preference to different items (i.e., N-to-N issue), which is pretty common in practice. Obviously, it is confusing for the TransE-like transition: the embeddings of items are close as long as a user who likes both of them are due to the some preference (Figure~\ref{fig:idea_e}), leading to an incorrect conclusion that a user consuming one shall consume the other, no matter what the user's preference is.

Such limitations are alleviated by introducing perference hyperplanes as illustrated in Figure~\ref{fig:idea_h}: $i$ and $i'$ have different representations, and are similar only when they are projected to a specific hyperplane. We thus define the hyperplane-based translation function as follows:

\begin{equation}
  g(u,i;p)=\parallel \mathbf{u}^{\bot} + \mathbf{p} - \mathbf{i}^{\bot} \parallel
\end{equation}

\noindent where $\mathbf{u}^{\bot}$ and $\mathbf{i}^{\bot}$ are projected vectors of the user and the item, and are obtained through the induced preference $p$ that plays a similar role as relations in TransH:

\begin{equation}
  \mathbf{u}^{\bot}=\mathbf{u}-\mathbf{w}_p^{\mathrm{T}} \mathbf{u} \mathbf{w}_p
\end{equation}

\begin{equation}
  \mathbf{i}^{\bot}=\mathbf{i}-\mathbf{w}_p^{\mathrm{T}} \mathbf{i} \mathbf{w}_p
\end{equation}

\noindent where $\mathbf{w}_p$ is the projection vector that is obtained along with the induction process of preferences $p$: either to pick up the corresponding one using the hard strategy, or through attentive addition of all projection vectors based on the induced attention weights in the soft strategy:

\begin{equation}
  \mathbf{w}_p=\sum_{p'\in\mathcal{P}}\alpha_{p'} \mathbf{w}_{p'}
\end{equation}

We encourage the translation distances of the interacted items to be smaller than random ones for each user through BPR Loss function:

\begin{equation}
  \mathcal{L}_p=\sum_{(u,i)\in\mathcal{Y}}\sum_{(u,i')\in\mathcal{Y}'}-\log\sigma[g(u,i';p')-g(u,i;p)]
\end{equation}

\noindent where $\mathcal{Y}'$ contains negative interactions by randomly corrupting an interacted item to a non-interacted one for each user.

Tranditional methods (e.g., BPRMF~\cite{rendle2009bpr}) recommend items for a user by computing a scalar score based on user and item embeddings, which indicates to which extent the user prefers to the item. Instead, we model preferences as vectors in order to (1) capture the commonality among users as global latent features, compared with user embeddings which only concerns with the local features of the user alone, and (2) reflect richer semantics for explainable ability.

\section{Joint Learning via KTUP for Two Tasks}
\label{sec:joint}

KTUP extends the translation-based recommendation model, TUP, by incorporating KG knowledge of entities as well as relations. Intuitively, the auxiliary knowledge supplements the connectivity among items as constraints to model user-item pairs. On the other hand, the understanding of users' preferences to items shall reveal their commonality related to some relation types and entities, which may be missing in the given KG.

\subsection{KTUP}

Figure~\ref{fig:framework} presents the overall framework of KTUP. On the left side is the inputs: user-item interactions, knowledge graph and the alignments between items and entities. At the top-right corner is TUP for item recommendation, while TransH for knowledge graph completion is at the bottom-right corner. KTUP jointly learns the two tasks by enhancing the embeddings of items and preferences with that of entities and relations. We define the knowledge enhanced TUP translation function as follows:

\begin{equation}
  g(u,i;p)=\parallel \mathbf{u}^{\bot} + \hat{\mathbf{p}} - \hat{\mathbf{i}}^{\bot} \parallel
  \label{eq:ktup}
\end{equation}

\noindent where $\hat{\mathbf{i}}^{\bot}$ is the projected vector for the enhanced item embedding $\hat{\mathbf{i}}$ by the corresponding entity embedding $\mathbf{e}$:

\begin{equation}
  \hat{\mathbf{i}}^{\bot}=\hat{\mathbf{i}}-\hat{\mathbf{w}}_p^{\mathrm{T}} \hat{\mathbf{i}} \hat{\mathbf{w}}_p
\end{equation}

\begin{equation}
  \hat{\mathbf{i}}=\mathbf{i}+\mathbf{e},\ (i,e)\in\mathcal{A}
\end{equation}

And $\hat{\mathbf{p}}$ and $\hat{\mathbf{w}}_p$ are the translation vector and the projection vector enhanced by those of the corresponding relation embedding according a predefined one-to-one mapping $\mathcal{R}\rightarrow\mathcal{P}$. We obtain these two vectors as follows:

\begin{equation}
  \hat{\mathbf{p}}=\mathbf{p}+\mathbf{r}
\end{equation}

\begin{equation}
  \hat{\mathbf{w}}_p=\mathbf{w}_p+\mathbf{w}_r
\end{equation}

Thus, for entities and items, the enhanced item embeddings contain the relational knowledge among items that is complementary to user-item interactions, and improves item recommendation, since the entity embedding $\mathbf{e}$ perserves the structural knowledge in KG. Meanwhile, the entity embedding $\mathbf{e}$ shall be fine tuned by the additional connectivity through users and items during backpropagation. Note that we don't use the combined embeddings for both tasks, because it makes the embeddings of items the same as corresponding entities in two tasks, which actually degrades our model to share embeddings between items and entities.

For relations and preferences, the usage of relations not only offers explicit interpretation of explainability, but also further combines the two tasks more sufficiently at model level. On one hand, through the one-to-one mapping, the meaning of each preference is revealed by the relation label. For instance, the relation \textit{isDirectorOf} reveals a preference to director, or \textit{starring} for a preference to movie stars. On the other hand, many items have no aligned entities due to the incompleteness of KG, which limits the mutual impacts to the alignments between entities and items in the models that only transfer knowledge of entities. Considering that each user-item pair has a preference and so does the relations between two entities, KTUP optimizes all users, items and entities more thoroughly.

\subsection{Training}
We train KTUP using the overall objective function as follows:

\begin{equation}
  \mathcal{L} = \lambda \mathcal{L}_p + (1-\lambda) \mathcal{L}_k
\end{equation}

\noindent where $\lambda$ is a hyperparameter to balance the two tasks.
  
\subsection{Relationship to SOTA Models}
In this section, we give a discussion on the relationship between KTUP and the other state-of-the-art KG-based recommendation methods to facilite a deep understanding between two tasks in Section~\ref{sec:exp}. We choose three typical models that transfer knowledge of entities at data level (CFKG~\cite{DBLP:journals/corr/abs-1803-06540}), at embedding level (CKE~\cite{CKE}) and in both directions (CoFM~\cite{DBLP:conf/esws/PiaoB18}). We summerize the main differences and similarities from the following aspects:

\para{Implicity of User Preference} CKE and CoFM can be regarded as extentions of collaborative filtering. This type of methods consider the preferences from users to items implicitly and rely on their embeddings to compute a score (i.e., dot product) indicating in which degree the user likes the item. CFKG and KTUP model the preferences explictly and learn the vectoral representations instead of scalars to capture more comprehensive semantics.

\para{Variety of User Preference} CFKG defines the only \textit{buy} preference between users and items, which obviously suffers from the serious N-to-N issue and fails to deal with it through the TransE-like scoring function. TKUP distinguishes different user perferences and introduces hyperplanes for each preference as well as each relation to learn the various representations of items and entities.

\para{Transfered Knowledge from KG} CKE and CoFM only focus on transfering knowledge of entities. CFKG also transfers relations in the way of data integration through a unifed graph. Except for entities and items, KTUP combines the embeddings of relations and preferences according to the predefined one-to-one mapping, which brings another byproduct of the explainable ability to recommendation mechanism.

\section{Experiments}
\label{sec:exp}
In this section, we conduct quantitative experiments on separate tasks of item recommendation and KG completion. For each task, we use two datasets in different domains and give further evaluations on the influence of data sparsity as well as the N-to-N issue. We then investigate the mutual impacts between the two tasks during joint training. Finally, we highlight real examples for qualitative analysis to give an intuitive impression of explainability. We publish our project at \url{https://github.com/TaoMiner/joint-kg-recommender}.

\subsection{Datasets}
Following CoFM~\cite{DBLP:conf/esws/PiaoB18}, we use two publicly available datasets in the movie and book domains: MovieLens-1m~\cite{noia2016sprank} and DBbook2014~\footnote{\url{http://2014.eswc-conferences.org/important-dates/call-RecSys.html}}. Both datasets consist of users and their ratings on movies or books, which are then refined for LODRecSys~\cite{heitmann2012open,noia2016sprank,piao2017factorization} by mapping items into DBPedia entities if there is a mapping available. Following most item recommendation work that models implicit feedback~\cite{xiang2018reasoning}, we treat existing ratings as positive interactions, and generate negative ones by randomly corrupting items.

To collect the related facts from DBPedia, we only consider those triplets that are directly related to the entities with mapped items, no matter which role (i.e. subject or object) the entity serves as. We then preprocess the two datasets by: filtering out low frequency users and items (i.e., lower than 10 in MovieLens and 5 in DBbook), filtering out infrequent entities (i.e., lower than 10 in both datasets), cutting off unrelated relations and merging similar relations manually.

\begin{table}[htbp]
  \centering
  \caption{Statistics of MovieLens-1m and DBbook2014}
    \label{tab:data}
  \begin{tabular}{c|c|c|c}
  \hline
   &  & \textbf{MovieLens-1m} & \textbf{DBbook2014} \\ \hline
  \multirow{5}{*}{\begin{tabular}[c]{@{}l@{}}User-Item\\ Interactions\end{tabular}} & \# Users & 6,040 & 5,576 \\ \cline{2-4} 
   & \# Items & 3,240 & 2,680 \\ \cline{2-4} 
   & \# Ratings & 998,539 & 65,961 \\ \cline{2-4} 
   & \# Avg. ratings & 165 & 12 \\ \cline{2-4} 
   & Sparsity & 94.9\% & 99.6\% \\ \hline
  \multirow{3}{*}{KG} & \# Entity & 14,708 & 13,882 \\ \cline{2-4} 
   & \# Relation & 20 & 13 \\ \cline{2-4}
   & \# Triple & 434,189 & 334,511 \\ \hline
   \multirow{2}{*}{Multi-Tasks} & \begin{tabular}[c]{@{}l@{}}\# Item-Entity\\ Alignments\end{tabular} & 2,934 & 2,534 \\ \cline{2-4}
   & Coverage & 90.6\% & 94.6\% \\ \hline
  \end{tabular}
  \end{table}

Table~\ref{tab:data} shows the statistics of MovieLens-1m and DBbook2014 datasets\footnote{Because we cannot find the released triplets for the two datasets, we collect them as our KG as mentioned above, which leads to a little difference with those papers using the same datasets (e.g., CoFM~\cite{DBLP:conf/esws/PiaoB18}).}. After preprocessing, there are 6,040 users and 3,230 items with 998,539 ratings in Movielens-1m, the average number of ratings per user is 165 and the sparisity rate is 94.9\%. The data sparisity issue is more severe in DBbook2014. It consists of 5,576 users and 2,680 items with 65,961 ratings, where the average number of ratings per user is 12 and the sparisity rate reaches up to 99.6\%. The triplets used in the two datasets are at the same scale, where the subgraph for MovieLens-1m composes of 434,189 triplets with 14,708 entities and 20 relations, while the subgraph of DBbook has 334,511 triplets with 13,882 entities and 13 relations. Note that the alignments between items and entities for transfering are fewer in MovieLens-1m than that in DBbook2014.

\subsection{Baselines}
For item recommendation, we compare our proposed models with the following state-of-the-art baselines involving typical similarity-based methods and KG-based methods.

\begin{itemize}
  \item Typical similarity-based methods: we choose the widely used collaborative filtering models, \textbf{FM}~\cite{rendle2010factorization} and \textbf{BPRMF}~\cite{rendle2009bpr}, because they are the foundations of other baselines and also achieve the state-of-the-art performance on many benchmark datasets.
  \item \textbf{CFKG}~\cite{DBLP:journals/corr/abs-1803-06540} that integrates the data of two sources and applies TransE on a unifed graph including users, items, entities and relations;
  \item \textbf{CKE}~\cite{CKE} that combines various item embeddings from different sources including TransR on KG;
  \item \textbf{CoFM}~\cite{DBLP:conf/esws/PiaoB18} that jointly trains FM and TransE by sharing parameters or regularization of aligned items and entities. We mark the two schemes as CoFM (share) and CoFM (reg), respectively.
\end{itemize}

For KG completion, we choose the typical methods \textbf{TransE}~\cite{bordes2013translating}, \textbf{TransH}~\cite{wang2014knowledge} and \textbf{TransR}~\cite{lin2015learning} that are widely used in this field. Also, we evaluate the above KG-based methods even if they have not been done so in the original papers to investigate the impacts of different transfer schemes.

For fair comparison, we carefully reimplement them in our released codes because they did not report the results on the same datasets and we cannot find their released codes. Note that we remove the components of side information modeling like reviews and visual information, since they are not available in the datasets and are out of the scope of this paper.

\subsection{Training Details}
We construct training set, validation set and test set by randomly spliting the dataset with the ratio of $7:1:2$. For item recommendation, we split the items for each user and ensure at least one item exist in the test set. 

For hyperparameters, we apply a grid search on BPRMF and TransE to find the best settings for each task, and use them for all of the other models since they share the basic learning ideas\footnote{We have tested other parameters for baselines and find performance drop.}. The learning rate is searched in $\{0.0005, 0.005, 0.001, 0.05, 0.01\}$, the coefficient of $L_{2}$ regularization is in $\{10^{-5},10^{-4},10^{-3},10^{-2},10^{-1},0\}$, and the optimization methods include Adaptive Moment Estimation (Adam), Adagrad and SGD. Finally, we set the learning rate as $0.005$ and $0.001$ for item recommendation and KG completion, respectively, $L_{2}$ coefficient is set to $10^{-5}$ and $0$, and the optimization methods is set to Adagrad and Adam. Particularly, for the models involving two tasks, we have tried both sets of parameters, and pick up the latter set of parameters due to its superior performance.

Other hypermeters are empirically set as follows: the batch size is $256$, the embedding size is $100$ and we perform early stopping strategy on the validation sets.

\begin{table*}[]
  \caption{Overall performance on Item Recommendation}
  \vspace{-0.2cm}
    \label{tab:itrec}
  \begin{tabular}{l|ccccc|lllll}
  \hline
   & \multicolumn{5}{c|}{\textbf{MovieLens-1m (@10, \%)}} & \multicolumn{5}{c}{\textbf{DBbook2014 (@10, \%)}} \\
   & \multicolumn{1}{l}{\textbf{Precision}} & \multicolumn{1}{l}{\textbf{Recall}} & \multicolumn{1}{l}{\textbf{F1}} & \multicolumn{1}{l}{\textbf{Hit}} & \multicolumn{1}{l|}{\textbf{NDCG}} & \textbf{Precision} & \textbf{Recall} & \textbf{F1} & \textbf{Hit} & \textbf{NDCG} \\ \hline
  FM & 29.28 & 11.92 & 13.81 & 81.06 & 59.48 & 3.44 & 21.55 & 5.75 & 30.15 & 20.10 \\
  BPRMF & 30.81 & 12.95 & 14.84 & 83.18 & 61.02 & 3.56 & 22.46 & 5.96 & 31.26 & 21.01 \\ \hline
  CFKG & 29.45 & 12.49 & 14.23 & 82.24 & 58.97 & 3.17 & 19.69 & 5.30 & 28.09 & 19.87 \\
  CKE & 38.67 & 16.65 & 18.94 & 88.36 & 67.05 & 3.92 & 23.41 & 6.51 & 33.18 & \textbf{27.78} \\
  CoFM (share) & 32.08 & 13.02 & 15.12 & 83.30 & 58.69 & 3.41 & 20.78 & 5.67 & 29.84 & 20.92 \\
  CoFM (reg) &  31.74 & 12.74 & 14.87 & 82.67 & 58.66 & 3.32 & 20.54 & 5.54 & 28.96 & 20.53 \\ \hline
  TUP (hard) & 37.29 & 17.07 & 18.98 & \textbf{89.60} & 67.40 & 3.40 & 21.11 & 5.67 & 29.56 & 20.19 \\
  TUP (soft) & 37.00 & 16.79 & 18.76 & 89.47 & 67.02 & 3.62 & 22.81 & 6.06 & 31.42 & 21.54 \\
  KTUP (hard) & 40.87 & 17.24 & 19.79 & 88.97 & 69.65 & 4.04 & 24.48 & 6.71 & 34.49 & 27.38 \\
  KTUP (soft) & \textbf{41.03} & \textbf{17.25} & \textbf{19.82} & 89.03 & \textbf{69.92} & \textbf{4.05} & \textbf{24.51} & \textbf{6.73} & \textbf{34.61} & 27.62 \\ \hline
  \end{tabular}
  \end{table*}

We predefine the number of preferences in TUP as $20$ and $13$ for MovieLens-1m and DBbook2014, respectively, which are set according to the relations of the collected triplets. For the models involving two tasks (i.e., CFKG, CKE, CoFM and KTUP), we set the joint hyperparameter $\lambda$ as $0.5$ and $0.7$ on the two datasets after searching in $\{0.7, 0.5, 0.3\}$, so as to balance their impacts, and use the pretrained embeddings of the basic model (i.e., BPRMF and TransE).

The main goal in this paper is to investigate the mutual impacts on each task during jointly training, rather than achieving best performance by tuning parameters. Thus, our proposed models as well as baseline methods are trained once for each dataset and evaluate for both tasks of item recommendation and KG completion.

\subsection{Item Recommendation}
In this section, we evaluate our models as well as the baseline methods on the task of item recommendation. Given a user, we take all items in test sets as candidates and rank them according to the scores computed based on the embeddings of users and items. Thus, the N items ranked at top are the recommended items.

\subsubsection{Metrics}

We use five evaluation metrics that have been widely used in previous work:

\begin{itemize}
  \item Precision@N: It is the fraction of the items recommended that are relevant to the user. We compute the mean of all users as the final precision. 
  \item Recall@N: It is the proportion of items relevant to the user that have been successfully recommended. We compute the mean of all users as the final recall.
  \item F1 score@N: It is the harmonic mean of precision at rank N and recall at rank N.
  \item Hit ratio@N: It is $1$ if any gold items are recommended within the top N items, otherwise $0$. We compute the mean of all users as the final hit ratio score. 
  \item nDCG@N: Normalized Discounted Cumulative Gain (nDCG) is a standard measure of ranking quality, considering the graded relevance among positive and negative items within the top $N$ of the ranking list.
\end{itemize}

\subsubsection{Overall Results}

Table~\ref{tab:itrec} shows the overall performance of our proposed models as well as the baseline methods, where \textit{hard} and \textit{soft} denote the two preference induction strategies in Section~\ref{sec:pi}. We can observe that:

\begin{itemize}
  \item Our proposed methods perform the best compared with the baseline methods on the two datasets. Particularly, TUP performs competitively to other KG-based models, while it doesn't require any additional information. This is because TUP automatically infers the knowledge of preferences from the user-item interactions, and performs much better especially when the amount of interaction data is sufficient, like MovieLens-1m. By incorporating KG, KTUP further presents more promising improvements on DBbook than MovieLens (i.e., 11.06\% v.s. 4.43\% gains in F1), which implies that the knowledge is more helpful for sparse data.
  \item The hard strategy performs better than the soft strategy only when it is used for TUP on MovieLens-1m, which implies that to induce a deterministic user perference requires sufficient data, and the soft strategy is more robust.
  \item CFKG and CoFM perform slightly better than the typical models (i.e., FM and BPRMF) on MovieLens-1m, but perform worse on the sparse dataset of DBbook2014. One possible reason is that they both transfer entities by forcing their embeddings to be similar with the aligned items, leading to the loss of knowledge that has been perserved in the embeddings, and the loss becomes more serious when there is insufficient training data.
  \item CKE achieves pretty good performance on the two datasets mainly because it combines the embeddings of items and entities that perserve the information from both sources, instead of aligning them with similar positions in the latent space.
  \item All models preform much better on MovieLens-1m than on DBbook2014 due to the relatively sufficient training data and an easier test (a higher value even using random initializations). Interestingly, the improvement by utilizing KG is larger on dense dataset of MovieLens than that on the sparse dataset of DBbook. This goes against our intuitions that the more sparse the dataset is, the more potentials it shall have in absorbing richer knowledge. Thus, we further split the test set according to different sparisity levels of training data, and investigate the impacts from KG on each subset in the next section.
\end{itemize}

\begin{figure}[htb]
  \centerline{\includegraphics[width=0.48\textwidth]{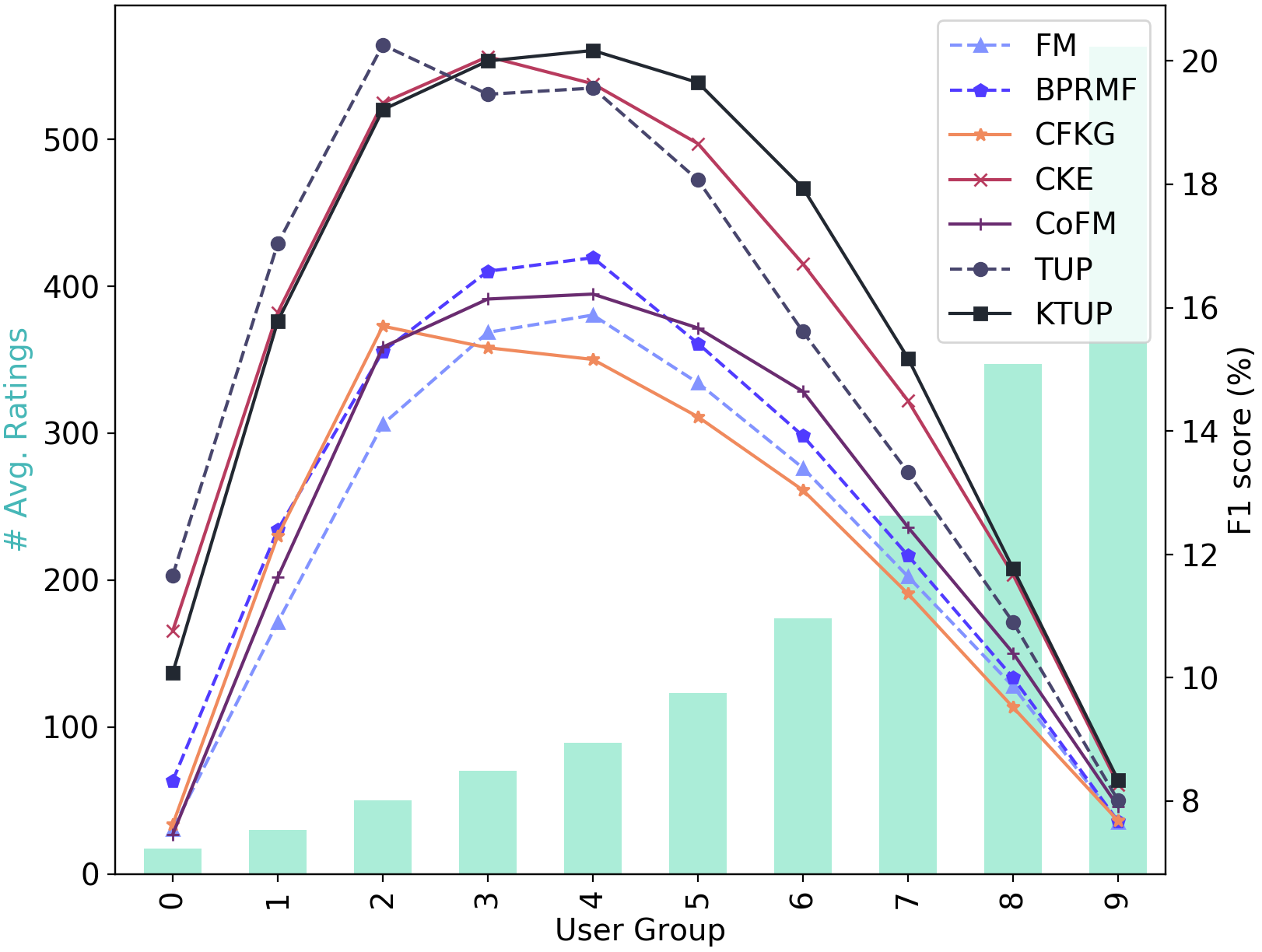}}
  \vspace{-0.2cm}
  \caption{Influence of Different Sparsity on MovieLens-1m. The x-axis shows 10 user groups splited according to interaction number, the left y-axis corresponds to the bars indicating the number of interactions in each user group, and the right y-axis denotes F1-score of curves.}
  \label{fig:rec_detail}
\end{figure}

\begin{table*}[htb]
  \caption{Performance on MovieLens by Relation Category}
  \vspace{-0.3cm}
  \label{tab:kgc_n2n_ml1m_hit}
  \begin{tabular}{c|cccc|cccc}
  \hline
  Task & \multicolumn{4}{c|}{\textbf{Prediction Head (Hits@10, \%)}} & \multicolumn{4}{c}{\textbf{Prediction Tail (Hits@10, \%)}} \\ \hline
  \textbf{Relation Category} & \textbf{1-to-1} & \textbf{1-to-N} & \textbf{N-to-1} & \textbf{N-to-N} & \textbf{1-to-1} & \textbf{1-to-N} & \textbf{N-to-1} & \textbf{N-to-N} \\ \hline
  TransE & 59.62 & 56.76 & 64.55 & 24.56 & \textbf{65.38} & 62.16 & 78.52 & 46.25 \\
  TransH & 61.54 & 48.65 & 65.73 & 25.51 & 57.69 & 78.38 & 75.62 & 46.73 \\
  TransR & 17.31 & 29.73 & 32.88 & 18.50 & 17.31 & 43.24 & 53.12 & 38.88 \\ \hline
  CFKG & 59.62 & 51.35 & 63.31 & 20.30 & 57.69 & 70.27 & 78.56 & 41.22 \\
  CKE & 19.23 & 21.62 & 24.16 & 14.81 & 7.69 & 24.32 & 37.83 & 34.82 \\
  CoFM (share) & 65.38 & 59.46 & 66.13 & 24.42 & 61.54 & 72.97 & \textbf{81.05} & 45.99 \\
  CoFM (reg) & 69.23 & \textbf{70.27} & 66.09 & 24.30 & 48.08 & \textbf{86.49} & 80.72 & 45.79 \\ \hline
  KTUP (hard) & 67.31 & 59.46 & 66.42 & 25.67 & 57.69 & 81.08 & 79.22 & 47.24 \\
  KTUP (soft) & \textbf{75.00} & 56.76 & \textbf{67.16} & \textbf{26.09} & 63.46 & 81.08 & 78.34 & \textbf{47.65} \\ \hline
  \end{tabular}
  \end{table*}

\subsubsection{Influence of Training Data Sparsity}

To investigate the influence of data sparsity on knowledge transfer, we split the test set of MovieLens-1m into 10 subsets according to the rating number of each user for training; meanwhile we also try to balance the number of users as well as ratings in each subset. The detailed results of F1 score are shown in Figure~\ref{fig:rec_detail}. Green bars represent the average rating number per user ranging from 17 to 563\footnote{Note that the sparisity of DBBook2014 can be concluded into the $0$th user group in MovieLens-1m, in which we observe similar results.}. We denote the models without KG knowledge as dashed lines, and other models as solid lines.

We can see that (1) KG-based methods (i.e., CKE and KTUP) outperform the other models the most when the average number of ratings per user ranges from 100 to 200. (2) The gap between the two types of models is getting closer as the amount of the training data decreases, and the improvements become similar with that on DBbook when their training data is at the similar sparisity level. (3) Meanwhile, the gap almost disappears when the average ratings are 563 (the left most bar), which implies that the impacts of KG become negligible if there if sufficient training data. Note that the performance of all models are getting worse when the average ratings are larger than 89. The possible reason is the user likes so many items that the preferences are too general to capture. (4) TUP outperforms KTUP when user preferences are relative simple to model (i.e., \#rating<50), showing the effectiveness and necessity to fully utilize user-item interactions for preference modeling.

\begin{table}[htb]
  \caption{Overall performance on KG Completion}
  \vspace{-0.2cm}
    \label{tab:kgc}
  \begin{tabular}{p{2cm}|cc|cc}
  \hline
   & \multicolumn{2}{c|}{\textbf{MovieLens-1m}} & \multicolumn{2}{c}{\textbf{DBbook2014}} \\
   & \begin{tabular}[c]{@{}c@{}}\textbf{Hit@10}\\ \textbf{(\%)}\end{tabular} & \begin{tabular}[c]{@{}c@{}}\textbf{Mean}\\ \textbf{Rank}\end{tabular} & \begin{tabular}[c]{@{}c@{}}\textbf{Hit@10}\\ \textbf{(\%)}\end{tabular} & \begin{tabular}[c]{@{}c@{}}\textbf{Mean}\\ \textbf{Rank}\end{tabular} \\ \hline
  TransE & 46.95 & 537 & 60.71 & 531 \\
  TransH & 47.63 & 537 & 60.06 & 556 \\
  TransR & 38.93 & 609 & 56.33 & 563 \\ \hline
  CFKG & 41.56 & 523 & 58.83 & 547 \\
  CKE & 34.37 & 585 & 54.66 & 593 \\
  CoFM (share) & 46.62 & 515 & 57.01 & 529 \\
  CoFM (reg) & 46.51 & \textbf{506} & 60.81 & 521 \\ \hline
  KTUP (hard) & 48.39 & 525 & 60.53 & 501 \\
  KTUP (soft) & \textbf{48.90} & 527 & \textbf{60.75} & \textbf{499} \\ \hline
  \end{tabular}
  \end{table}

\subsection{Knowledge Graph Completion}
In this section, we evaluate on the task of KG completion. It is to predict the missing entity $e_h$ or $e_t$ for a given triplet $(e_h,e_t, r)$. For each missing entity, we take all entities as candidates and rank them according to the scores computed based on entity and relation embeddings.

\subsubsection{Metrics}
We use two evaluation metrics that have been widely used in previous work~\cite{wang2014knowledge}:

\begin{itemize}
  \item Hit ratio@N: It is $1$ if the miss entity are ranked within the top N candidates, otherwise $0$. We compute the mean of all triplets as the final hit ratio score. 
  \item Mean Rank : It is the averaged rank of the missing entities, the smaller the better.
\end{itemize}

\subsubsection{Overall Results}

Table~\ref{tab:kgc} shows the overall performance. We can see that KTUP almost outperforms all the other models on both datasets except the mean rank value on MovieLens-1m. We argure this metric is less important since it is easily reduced by an obstinate triple with a low rank~\cite{wang2016text}. Compared with TransH, it achieves a larger improvement on Hit Ratio for MovieLens-1m as compared to that for DBbook2014 (2.67\% v.s. 1.15\%), because Movielens-1m contains more connectivities between users and items that are helpful for modeling structural knowledge between entities. We also observe that CFKG, CKE and CoFM show a performance drop as compared to their basic KG components: TransE and TransR. One reason may be that these methods force the embeddings of aligned entities to satisfy the other task of item recommendation, while the aligned entities are only a small portion (i.e. 19.95\% and 18.25\% on the two datasets), which actually degrades the learning for KG completion. Another reason is that the N-to-N issues of user preferences have negative impacts on the representation learning of entities and relations, especially for the \textit{buy} relation in CFKG. CKE takes this issue into account but TransR contains a lot of trainable parameters and does not work well on such a small training set.

\begin{figure*}[htb]
  \centering
  \subfigure[KTUP ($\rho=0.97$)]{
  \label{fig:jtransup_mf1}\includegraphics[width=0.24\textwidth]{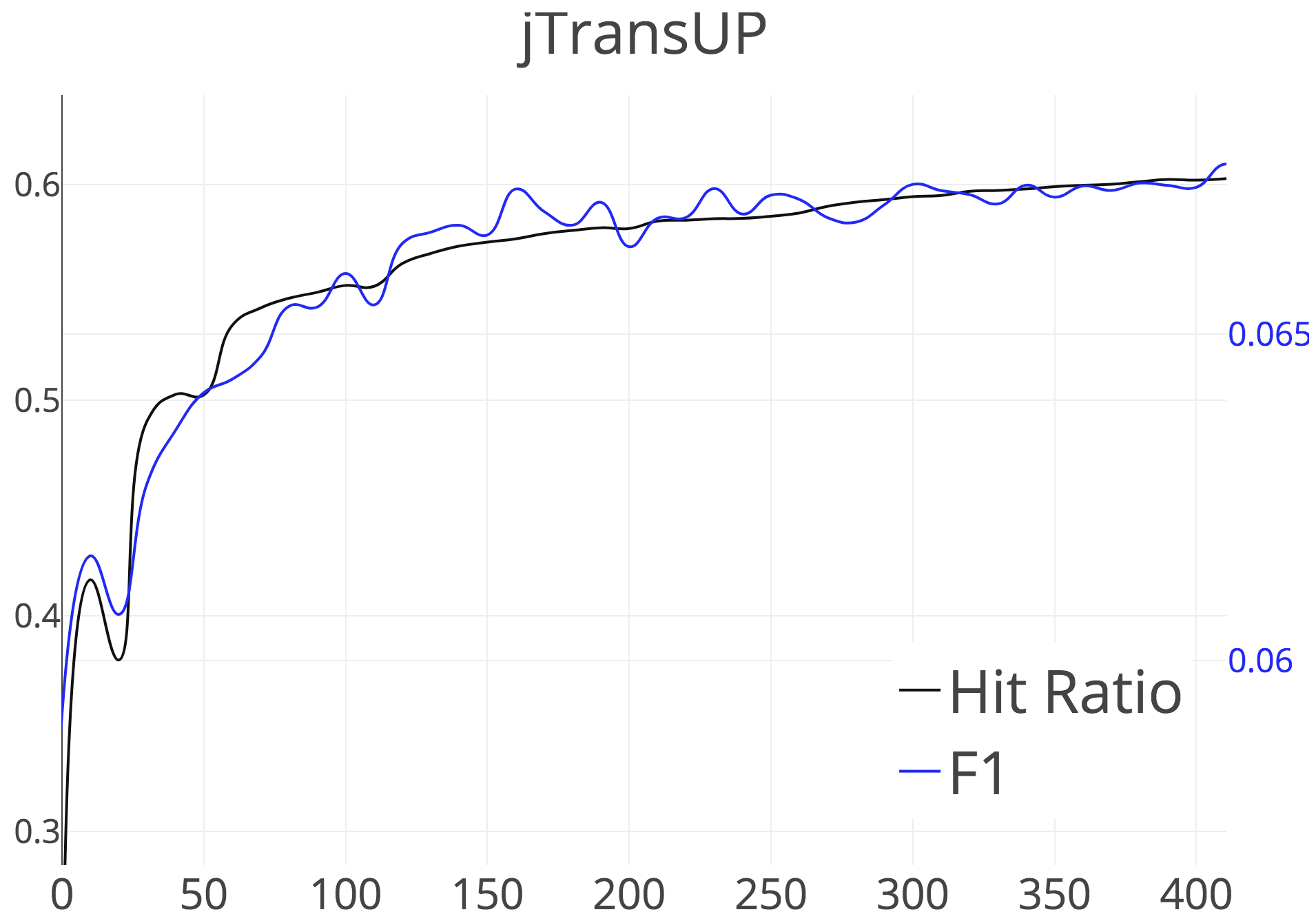}}
  \subfigure[CoFM ($\rho=0.81$)]{
  \label{fig:mhits}\includegraphics[width=0.24\textwidth]{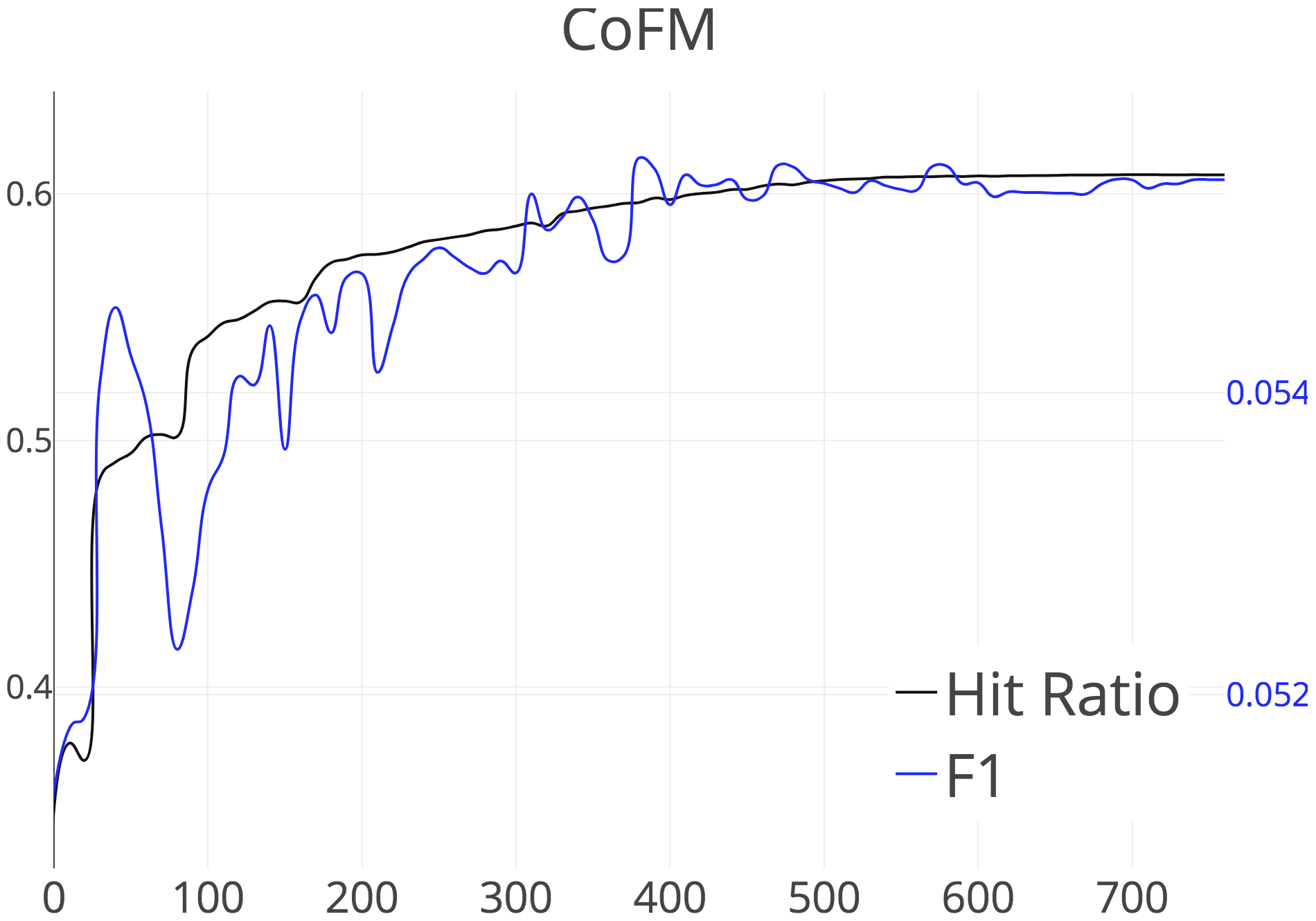}}
  \subfigure[CFKG ($\rho=0.97$)]{
  \label{fig:df1}\includegraphics[width=0.24\textwidth]{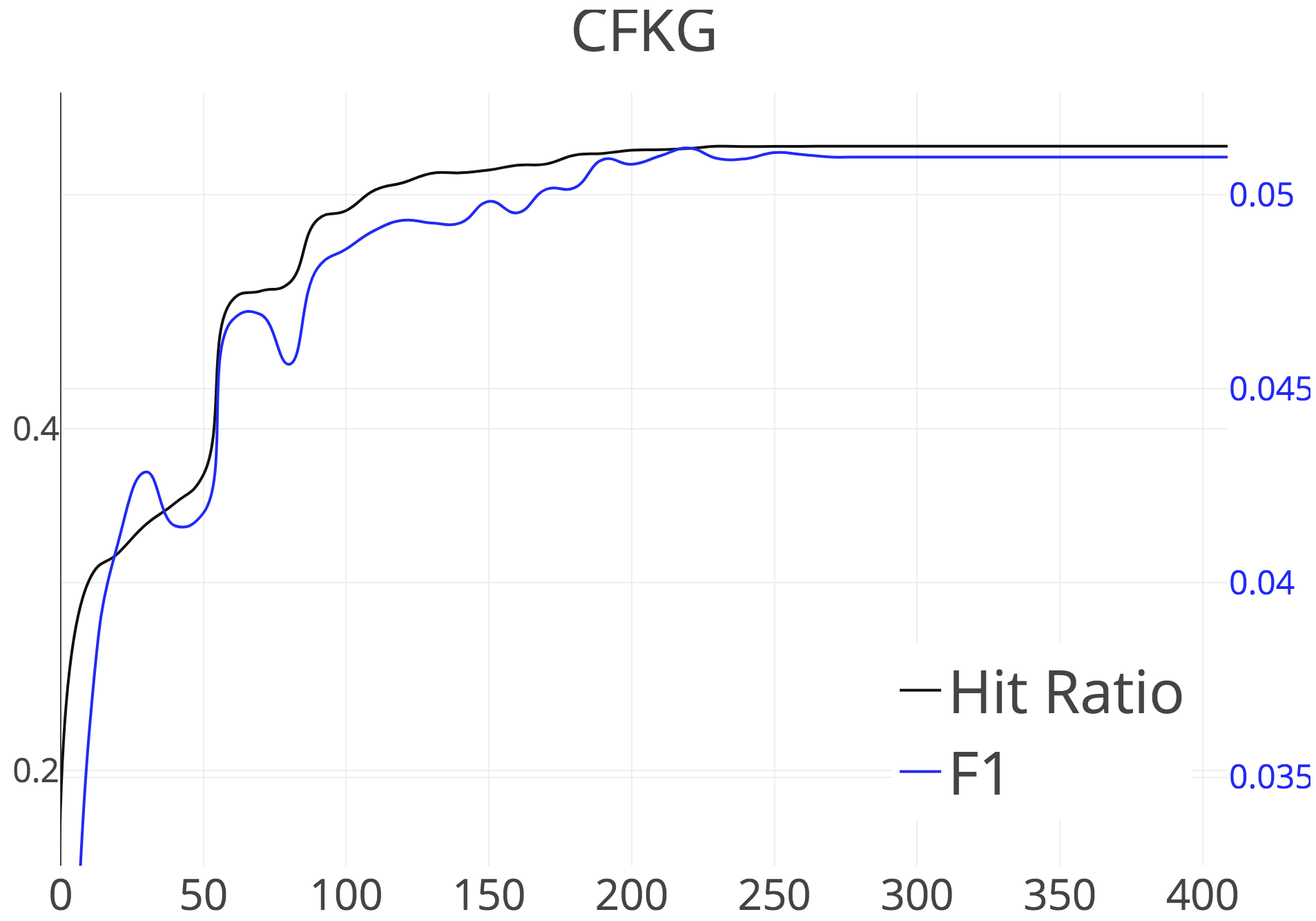}}
  \subfigure[CKE ($\rho=0.94$)]{
  \label{fig:dhits}\includegraphics[width=0.24\textwidth]{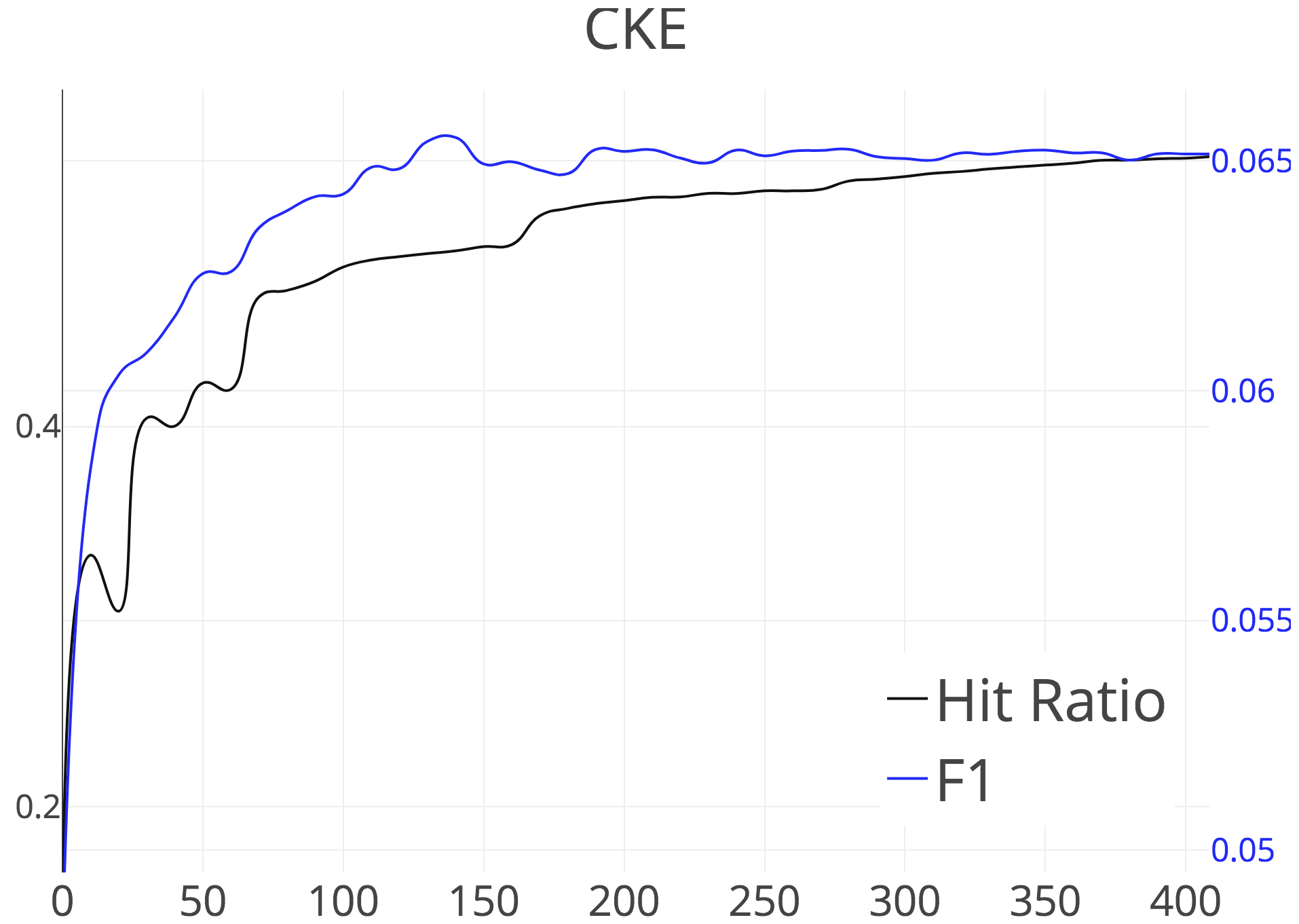}}
  \vspace{-0.3cm}
  \caption{Correlation of Training Curves between Two Tasks on DBbook2014, which is denoted by the Pearson's correlation coefficient $\rho$. The x-axis is training epoch, the left y-axis corresponds to KG completion via hit ratio, and the right y-axis is for item recommendation through F1. (Note that we scale the values of both F1 and Hit Ratio to the same magnitude.)}
  \label{fig:training_dbbook}
\end{figure*}

\begin{figure}[htb]
  \centerline{\includegraphics[width=0.48\textwidth]{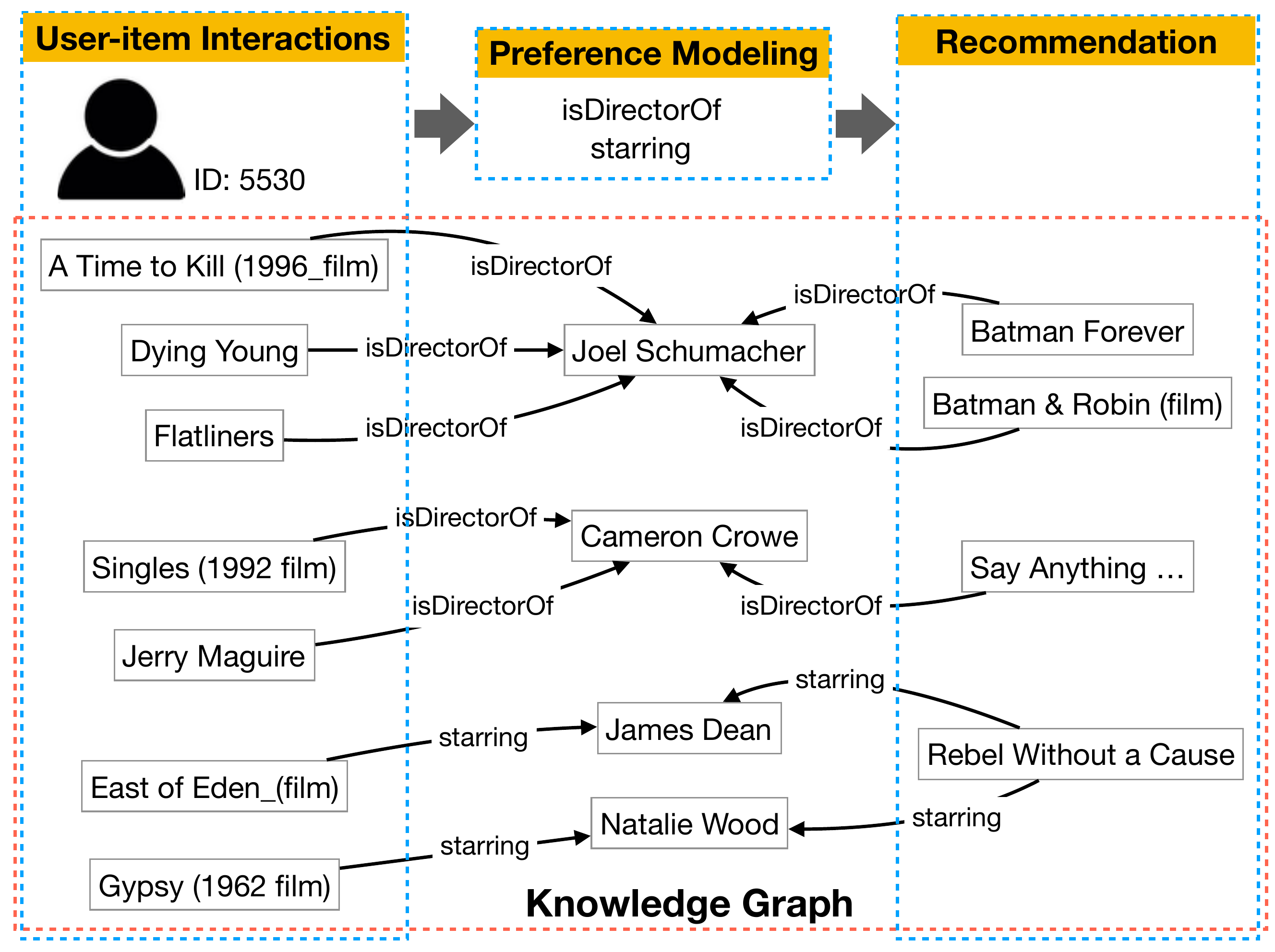}}
  \vspace{-0.3cm}
  \caption{Real Example from MovieLens-1m}
  \label{fig:qa}
\end{figure}

\subsubsection{Capability to Handle N-to-N Relations}
Table~\ref{tab:kgc_n2n_ml1m_hit} shows the separate evaluation results on each relation category. Following~\cite{bordes2013translating}, we divide relations into four types: 1-to-1, 1-to-N, N-to-1 and N-to-N. We can see that (1) TransR and its related models (i.e., CKE) perform the worst which is consistent with the above overall performance. (2) KTUP achieves the best performance on N-to-N issues, and also performs competitively with TransE and CoFM on 1-to-1, 1-to-N and N-to-1 problems, which indicates the capability of our methods in hanling complex relations and improves both tasks. (3) CFKG presents a lower value on N-to-N relations than TransE, which implies that the unifed graph may have introduced more confusing relational semantics. (4) CoFM performs competitively in KG completion task while worse in item recommendation, becuase their knowledge transfering schemes lead to unstable joint training. That is, it is difficult to control the positive impacts of knowledge transfer on which task, and different parameters for separately training CoFM on each task is required, which is also concluded in the original paper~\cite{DBLP:conf/esws/PiaoB18}.

\subsection{Mutual Benefits of Two Tasks}
Although the evalutions on separate tasks have been conducted, it is still unclear how different transfer schemes take effect. We thus investigate the correlations between the training curves of two tasks. Intuitively, a strong correlation implies a more complete transfer learning, and a better utilization of the complementary information from each other. Because KG completion has no F1 measures, so we present its hit ratio corresponds to left y-axis, and item recommendation through F1 is presented on right y-axis. 

As shown in Figure~\ref{fig:training_dbbook}, we can see that KTUP and CFKG present the strongest correlations between their curves, that is, the increase and decrease of one curve shall be reflected on the other curve simultaneously. This implies that the transfer of relations plays an important role in training the two tasks together. However, CFKG does not perform well on both tasks (as shown in Table~\ref{tab:itrec} and Table~\ref{tab:kgc}) mainly because of 2 reasons. First, it cannot deal with complex relations; second, it only increase the connectivity in the unified graph through the integration of relations and preferences, which are actually not transitional. Instead, KTUP combines the embeddings of relations and preferences that perserve two types of structural knowledge, and meanwhile introduces hyperplanes for the N-to-N issue. The curves of CoFM and CKE are obviously not correlated strongly due to the small portion of transfered entities. Concretely, CoFM forces the embeddings of aligned entities and items to be similar which may result in unstable training. CKE focuses on unidirectional enhancements by combining their embeddings, and thus performs well in item recommendation but worse in KG completion.

\subsection{Case Study}
In this section, we present an example of Movielens-1m to give an intuitive impression of our explainability. On the left is a user who has interacted with 7 movies. KTUP first induces user preferences to these movies, and finds that what the user is concerned is the relations of \textit{isDirectorOf} and \textit{starring} (the preferences having highest attention in Section~\ref{sec:pi}). Thus, it searches the nearest items according to Equation~\ref{eq:ktup} based on the induced preferences. We present the recommended four movies on the right side. In particular, \textit{Batman Forever} and \textit{Batman \& Robin (film)} are recommended because the user shows preference with their director \textit{Joel Schumacher}. Similarly, the preference to director also helps to induce the movie \textit{Say Anything ...} directed by \textit{Cameron Crowe}. Besides, the user also has preferences on \textit{starring}, such as \textit{James Dean} in \textit{East of Eden (film)} and \textit{Natalie Wood} in \textit{Gypsy (1962 film)}; together, the system suggests another movie \textit{Rebel Without a Cause}.

\section{Conclusion}

In this paper, we proposed a novel translation-based recommender model, TUP, and extended it to integrate KG completion seamlessly, namely KTUP. TUP is capable of modeling various implicit relations between users and items which reveal the preferences of users on consuming items. KTUP further enhances the model explainability via aligned relations and perferences, and improves the preformances of both tasks via joint learning.

In future, we are interested in inducing more complex user preferences over multi-hop entity relations and introducing KG reasoning (e.g., rule mining) techniques for unseen user preferences to deal with the cold start problem.

\begin{acks}
This research is part of NExT research which is supported by the National Research Foundation, Prime Minister's Office, Singapore under its IRC@SG Funding Initiative.
\end{acks}

\bibliographystyle{ACM-Reference-Format}
\balance 
\bibliography{ms}

\end{document}